\documentclass[12pt,preprint]{aastex}
                                                                                
\shorttitle{Constraints on the DGP Model from Recent Supernova
Observations and Baryon Acoustic Oscillations}	
\shortauthors{Guo, Z.-K., Zhu, Z.-H., Alcaniz, J.S. \& Zhang, Y.-Z.}

\begin{document}

\title{Constraints on the DGP Model from Recent Supernova
Observations and Baryon Acoustic Oscillations}
\slugcomment{\today}
\author{
Zong-Kuan Guo \altaffilmark{1},
Zong-Hong Zhu \altaffilmark{2},
J.S. Alcaniz \altaffilmark{3},
Yuan-Zhong Zhang \altaffilmark{4,1}
}
\altaffiltext{1}{Institute of Theoretical Physics, 
 Chinese Academy of Sciences, P.O. Box 2735, Beijing 100080, China}
\altaffiltext{2}{Department of Astronomy,
 Beijing Normal University, Beijing 100875, China}
\altaffiltext{3}{Departamento de Astronomia,
 Observat\'orio Nacional, 20921-400 Rio de Janeiro, Brasil}
\altaffiltext{4}{CCAST (World Lab.), P.O. Box 8730, Beijing 100080, China}


\begin{abstract}
Although there is mounting observational evidence that the expansion of our universe is undergoing a late-time acceleration, 
the mechanism for this acceleration is yet unknown. In the so-called Dvali-Gabadadze-Porrati (DGP) model this phenomena is attributed 
to gravitational \emph{leakage} into extra dimensions. In this work, we mainly focus our attention to the constraints on the model from 
the \emph{gold} sample of type Ia supernovae (SNeIa), the first year data from the Supernova Legacy Survey (SNLS) and the baryon acoustic oscillation (BAO) peak found in the Sloan Digital Sky Survey (SDSS). 
At 99.73\% confidence level, the combination of the three databases provides 
 $\Omega_m=0.270^{+0.018}_{-0.017}$ and $\Omega_{r_c}=0.216^{+0.012}_{-0.013}$
 (hence a spatially closed universe with $\Omega_k=-0.350^{+0.080}_{-0.083}$), 
 which seems to be in contradiction with the most recent WMAP results indicating a flat universe. 
Based on this result, we also estimated the transition redshift (at which the universe switches from deceleration 
to acceleration) to be $0.70 < z_{q=0} < 1.01$, at $2\sigma$ confidence level. 
\end{abstract}


%

\section{Introduction}

Recent observations of type Ia supernovae (SNe Ia) suggest that the expansion of the universe is accelerating 
(Riess et al. 1998, Perlmutter et al. 1999, Tonry et al. 2003, Barris et al. 2004, Knop et al. 2003, Riess et al. 2004). 
As is well known all usual  types of matter with positive pressure generate attractive forces, which decelerate the expansion 
of the universe. Given this, a dark energy component with negative pressure was suggested to account for the invisible fuel that drives 
the current acceleration of the universe. 
There are a huge number of candidates for the dark energy
  component in the literature (see, e.g., Sahni and Starobinsky 2000;
  Peebles and Ratra 2003; Padmanabhan 2003; Lima 2004;
  Copeland et al. 2006 for recent reviews),
  such as
  a cosmological constant $\Lambda$ (Carroll et al. 1992),
  an evolving scalar field
        (referred to by some as quintessence:
        Ratra and Peebles 1988;
        Caldwell et al. 1998;
	Weller and Albrech 2002;
	Guo, Ohta and Zhang 2005),
  the phantom energy, in which the sum of the pressure and energy
    density is negative
        (Caldwell 2002;
        Dabrowski et al. 2003;
	Wu and Yu 2005a),
   the quintom model
        (Feng, Wang and Zhang 2005;
        Guo et al. 2005;
        Zhao et al. 2005;
        Wu and Yu 2005b),
   the holographic dark energy
        (Li 2004;
        Gong 2004;
        Wang, Gong and  Abdalla 2005;
        Myung 2005;
        Zhang and Wu 2005;
        Pavon and Zimdahl 2005;
        Chang, Wu and Zhang 2006),
  the Chaplygin gas
        (Kamenshchik et al. 2001;
        Bento et al. 2002;
	Dev, Alcaniz and Jain 2003;
        Silva and Bertolami 2003;
        Makler et al. 2003;
        Zhu 2004;
        Gong 2005;
        Zhang and Zhu 2006),
  and the Cardassion model
        (Freese and Lewis 2002;
        Zhu and Fujimoto 2002, 2003;
	Godlowski, Szydlowski and Krawiec 2004;
        Amarzguioui, Elgaroy and Multamaki 2005;
        Koivisto, Kurki-Suonio and  Ravndal 2005;
        Lazkoz and Leon 2005;
        Szydlowski and Godlowski 2006).

Another possible explanation for the accelerating expansion of the universe could be the infrared modification of gravity expected from 
extra dimensional physics, which would lead to a modification of the effective Friedmann equation at late times. An interesting model 
incorporating modification of gravitational laws at large distances was proposed by Dvali, Gabadadze and Porrati (2000), the so-called DGP model. 
It describes our four-dimensional world as a brane embedded into flat five-dimensional bulk. While ordinary matter fields are supposed to be localized 
on the brane gravity can propagate into the bulk.  Unlike popular braneworld theories at the time, the extra dimension featured in this theory is 
astrophysically large and flat (for a recent review of the DGP phenomenology, see Lue 2005). A crucial ingredient of the model is the induced 
Einstein-Hilbert action on the brane. In this model, gravitational leakage into the bulk leads to the observed 
late-time accelerated expansion of the universe.
Such a possible mechanism for cosmic acceleration has been tested in many 
  of its observational predictions, ranging from local gravity
	(Lue 2003; Lue and Starkman 2003; Lue, Scoccimarro and Starkman 2004)
  to cosmological observations, 
	such as SNe Ia (Deffayet, Dvali and Gabadadze 2002; 
		Deffayet et al. 2002; Avelino and Martins 2002;
		Dabrowski et al. 2004;
        	Alam and Sahni 2005;
        	Maartens and Majerotto 2006),
	angular size of compact ratio sources (Alcaniz 2002),
	the age measurements of high redshift objects 
		(Alcaniz, Jain and Dev 2002),
	the optical gravitational lensing surveys (Jain, Dev and Alcaniz 2002),
	the large scale structures (Multam\"aki et al. 2003),
	and the X-ray gas mass fraction in galaxy clusters 
		(Zhu and Alcaniz 2005; Alcaniz and Zhu 2005).

This paper aims at placing new observational constrains on the DGP model by
  using the gold sample of 157 SNe Ia compiled by Riess et al. (2004), the
  71 new SNe Ia released recently by the Supernova Legacy Survey (SNLS)
  (Astier et al. 2005), and the baryon acoustic oscillations detected in the 
  large-scale correlation function of Sloan Digital Sky Survey (SDSS) 
  luminous red galaxies (Eisenstein et al. 2005).
It is shown that, if only the SNe Ia databases are used, $\Omega_{r_c}$ 
  and $\Omega_m$ are highly degenerated.
However when we combine the baryon acoustic oscillations found by 
  Eisenstein et al. (2005) from the SDSS data for analyzing, the degeneracy 
  between $\Omega_{r_c}$ and $\Omega_m$ is broken and the two parameters are
  accurately determined.

We stuctured this paper as follows. Section~2 discusses the basic expressions of the DGP model. In Section~3, we present our 
analysis of the model using the updated SNe Ia data and the baryon acoustic oscillations found in the SDSS data. We end the paper by 
discussing its main results in Section~4.

                                                                                
\section{Basic expressions of the DGP model}

In the DGP model the modified Friedmann equation due to the
  presence of an infinite-volume extra dimension reads
  (Deffayet, Dvali and Gabadadze 2002; Deffayet et al. 2002)
\begin{equation}
\label{eq:ansatz}
H^2 = H_0^2 \left[ 
	\Omega_k(1+z)^2+\left(\sqrt{\Omega_{r_c}}+ 
	\sqrt{\Omega_{r_c}+\Omega_m (1+z)^3}\right)^2
		\right]
\end{equation}
where $H$ is the Hubble parameter ($H_0$ is its current value), $\Omega_k$ and $\Omega_m$ represent the fractional contribution of curvature 
and of the matter (both baryonic and nonbaryonic), respectively, and $\Omega_{r_c}$,  the bulk-induced term, is defined as 
\begin{equation}
\label{eq:omegarc}
\Omega_{r_c} \equiv 1/4r_c^2H_0^2.
\end{equation}
In the above equatios, $r_c$  is the crossover scale beyond which the gravitational force follows the 5-dimensional $1/r^3$ behavior. Note that 
on short length scales $r \ll r_c$ (at early times) the
  gravitational force follows the usual four-dimensional
  $1/r^2$ behavior, i.e., the standard cosmological models
  are recovered. It has been shown that by
  setting the crossover scale $r_c$ close to the horizon size,
  this extra contribution to the Friedmann equation leads to
  acceleration which can in principle explain the supernova
  data (Deffayet, Dvali and Gabadadze 2002; Deffayet et al. 2002).
From Eq.~1  we find that the normalization condition is given by $\Omega_k + (\sqrt{\Omega_{r_c}}+ \sqrt{\Omega_{r_c} + \Omega_m}\,)^2 = 1$, 
while for a spatially flat scenario it reduces to $\Omega_{r_c}=(1-\Omega_m)^2/4$.

The current value of the deceleration parameter, defined
$q \equiv -a\ddot{a}/\dot{a}^2$, takes the form (Zhu and Fujimoto 2003, 2004)
\begin{equation}
q_0 = \frac{3}{2}\Omega_m\left(1
 +\frac{\sqrt{\Omega_{r_c}}}{\sqrt{\Omega_{r_c}+\Omega_m}}\right)
 -\left(\sqrt{\Omega_{r_c}}+\sqrt{\Omega_{r_c}+\Omega_m}\right)^2.
\end{equation}
The transition redshift $z_{q=0}$ at which the universe
switches from deceleration to acceleration, can be expressed
in the following analytic form (Zhu and Alcaniz 2005)
\begin{equation}
z_{q=0} = -1 + 2\left(\frac{\Omega_{r_c}}{\Omega_m}\right)^{1/3}.
\end{equation}

Note that from a phenomenological standpoint, the DGP model is a testable scenario with the same number of parameters as the $\Lambda$CDM 
scenario, contrasting with models of quintessence that have additional free parameters to be determined 
(Deffayet et al. 2002).


\section{Constraints from SNeIa and SDSS data}

In this section we analyze the DGP model by using two
  recently released supernova data sets, the Gold supernova
  data set (Riess et al. 2004) and the SNLS data set (Astier et al. 2005).
We also use these data sets in conjunction with the recent
  discovery of the baryon acoustic oscillation peak in the
  SDSS (Eisenstein 2005) to place constrains on the cosmological parameters.

Recently, Riess et al. (2004) compiled a large database
  of 170 previously reported SNe Ia and 16 new high redshift
  SNe Ia observed by the Hubble Space Telescope (HST). 
The total sample span a wide range of redshift ($0.01 < z < 1.7$).
To reflect the difference in the quality of the spectroscopic
  and photometric record for individual supernovae, they
  divided the total sample into ``high-confidence" (gold) and
  ``likely but not certain" (silver) subsets. Here, we consider
  only the gold sample of 157 SNe Ia (for recent usages of the sample,
  see, e.g., Padmanabhan and Choudhury 2003; 
	Nesseris and Perivolaropoulos 2004; Alcaniz 2004; Choudhury and Padmanabhan 2005;
	Gong 2005;
	Feng, Wang and Zhang 2005;
	Zhang and Wu 2005;
	Guo and Zhang 2005a,b;
	Cai, Gong and Wang 2006;
        Ichikawa and Takahashi 2005).

More recently, the SNLS collaboration released the first year data of its planned five-year Supernova Legacy Survey (Astier et al. 2005). 
An important aspect to be emphasized on the SNLS data is that they seem to be in a better agreement with WMAP results than the gold sample 
(see, e.g., Jassal, Bagla and Padmanabhan 2006). The two samples are illustrated on a residual Hubble Diagram
  with respect to our best fit universe
 ($\Omega_m = 0.270$, $\Omega_{r_c} = 0.216$)
  in Figure~1.

It is well know that the acoustic peaks in the cosmic microwave background 
  (CMB) anisotropy power spectrum can be used to determine the properties of
  the cosmic perturbations, to measure the contents and curvature of the 
  universe, as well as many other cosmological parameters 
  (see, e.g., Spergel et al. 2003). 
Because the acoustic oscillations in the relativistic plasma of the early
  universe will also be imprinted on to the late-time power spectrum of the
  non-relativistic matter (Peebles and Yu 1970; Eisenstein and Hu 1998),
  the acoustic signatures in the large-scale clustering of galaxies yield
  additional tests for cosmology.
In particular, the characteristic and reasonably sharp length scale measured
  at a wide range of redshifts provides distance-redshift relation, which is
  a geometric complement to the usual luminosity-distance from type Ia 
  supernove (Eisenstein et al. 2005). 
Although the acoustic features in the matter correlations are weak and on 
  large scales, Eisenstein et al. (2005) have successfully found the peaks
  using a large spectroscopic sample of luminous, red galaxies (LRGs) from
  the Sloan Digital Sky Sruvey (SDSS, York et al. 2000).
This sample contains 46,748 galaxies covering 3816 square degrees out to a
  redshift of z=0.47.
They found a parameter A, which is independent of dark energy models
  (Eisenstein et al. 2005).
From their Eq.~2 and 4, we write it as follows,
\begin{equation}
A = \frac{\sqrt{\Omega_{m}}}{z_1}
 \left[\frac{z_1}{E(z_1)}\frac{1}{|\Omega_k|} {\rm sinn}^2
 \left(\sqrt{|\Omega_k|}\int_0^{z_1}\frac{dz}{E(z)}\right)\right]^{1/3},
\end{equation}
where $E(z) \equiv H(z)/H_0$, $z_1 = 0.35$, $A$ is measured
to be $A = 0.469 \pm 0.017$, and the function ${\rm sinn}(x)$
is defined as ${\rm sinn}(x) = \sin(x)$ for a closed
universe, ${\rm sinn}(x) = \sinh(x)$ for an open universe
and ${\rm sinn}(x) = x$ for a flat universe.
In our analysis, we will comebine these measuremets.
                                                                                
%
%

In order to place limits on our Eq. (1), we perform a $\chi^2$-statistics for the model parameters ($\Omega_m$, $\Omega_{r_c}$) and 
the Hubble constant $H_0$. Since we want to concentrate solely on the density parameters, we need to marginalize over the Hubble parameter 
$H_0$. However, $H_0$ appears as a quadratic term in $\chi^2$ or, equivalently, appears as a separable
  Gaussian factor in the probability to be marginalized over.
Thus marginalizing over $H_0$ is equivalent to evaluating $\chi^2$ at its
  minimum with respect to $H_0$ (Barris et al. 2004). 
Here, we marginalize over the Hubble parameter by using the
  analytical method of Wang et al. (2004).
Figure~2 shows the joint confidence contour at 68.3\%, 95.4\% and 99.7\%
  confidence levels in the parametric space $\Omega_m-\Omega_{r_c}$ arising  from the gold sample of SN Ia data and the
  SDSS baryon acoustic oscillations.
The best-fit parameters for this analysis are $\Omega_m = 0.272$ and $\Omega_{r_c} = 0.211$.
Note that the best-fit value for $\Omega_{r_c}$ leads to an
  estimate of the crossover scale $r_c$ in terms of the Hubble
  radius $H^{-1}_0$, i.e., $r_c = 1.089H^{-1}_0$.
Compared to Figure 2 of Alcaniz and Pires (2004), the model
  parameters are more tightly
  constrained by using the prior from the baryon oscillation
  results than by assuming a Gaussian prior on the matter
  density parameter, $\Omega_m = 0.27 \pm 0.04$, as
  provided by WMAP team (Spergel et al. 2003).

%
%

Figure~3 illustrates the allowed regions in the $\Omega_m-\Omega_{r_c}$ plane by using the first year SNLS data in conjunction with the SDSS 
baryon acoustic oscillations (see also Fairbairn and Goobar (2005) for a similar analysis\footnote{During the writing of this work we bacame 
aware of the results of Fairbairn and Goobar (2005). In their analysis, however, they paid particularly more attention to a generalized verison of 
DGP model.}). Our best-fit for this joint SNLS plus BAO analysis happens at $\Omega_m = 0.265$ and $\Omega_{r_c} = 0.216$. The parameter space is 
considerably reduced relative to Figure~2 since the SNLS data set is more sensitive to the value of $\Omega_{r_c}$ than the gold sample.

%
%

In Figure~4 we show the joint confidence contours from the gold sample of SN Ia data and the first year SNLS data. In this case, the best-fit 
model happens for $\Omega_m = 0.31$ and $\Omega_{r_c} = 0.23$.
We find that the degeneracies between these parameters are broken by 
  combining these two data sets in the joint statistical analysis.
With the prior from the SDSS baryon acoustic oscillations, our fits provide
  $\Omega_m = 0.270$ and $\Omega_{r_c} = 0.216$.
Compared to Figure~4, the allowed confidence regions are slightly reduced.

%
%

Note that a closed universe is obtained at $3\sigma$ confidence level
  in the above analyses, which confirms the previous results obtained 
  using the SNe Ia and the X-ray mass fraction data of galaxy clusters
  (Zhu and Alcaniz 2005; Alcaniz and Zhu 2005).
Although there is a range on the parameter plane being consistent with both the SNeIa and the SDSS data, and the resulting matter density $\Omega_m$ 
is reasonable, a closed universe is obtained at a 99\% confidence level, which seems to be inconsistent with the result, 
$\Omega_k=-0.02^{+0.02}_{-0.02}$, found by the WMAP team (Bennett et al. 2003, Spergel et al. 2003) and $\Omega_k=0$ predicted by the 
simplest inflationary scenarios. Avelino and Martins (2002) analyzed the same model with the 92 SNe Ia from Riess et al. (1998) and 
Perlmutter et al. (1999). Assuming a flat universe, the authors obtained a low matter density and claimed the model was disfavorable. 
{\bf{In additional to including new SN Ia data from, and combining the 
  SDSS data, we relax the flat universe constraint in our analysis.}} 
We obtained a reasonable matter density, but a closed universe.
This means that, in light of WMAP results -- a nearly flat universe with $\Omega_k=-0.02^{+0.02}_{-0.02}$ --  the accelerating universe 
from gravitational leakage into an extra dimension seems not to be favored by the current observational data.
Note also that the best fit values of $\Omega_m$ and $\Omega_{r_c}$ 
  lead to an estimate of the transition redshift
$z_{q=0} = 0.86^{+0.07}_{-0.08}$, which is larger than the
  one estimated from the gold sample, i.e.,
  $z_{q=0} = 0.46 \pm  0.13$ (Riess et al. 2004).
It means that acceleration in the DGP model happens earlier.
Figure~5 shows the deceleration parameter as a function of redshift
  $z$ for our best-fit values in DGP model.
For comparison, we have also plot the curve for the standard $\Lambda$CDM
  model.
In Table~1 we summarize the main results of the paper.
%
%
\begin{table}[t]
\begin{center}
Table 1: Constrains on $\Omega_m$, $\Omega_{r_c}$, $r_c$ and $z_{q=0}$
{\footnotesize
\begin{tabular}{l l l l l} \hline\hline
Test & $\Omega_m$ & $\Omega_{r_c}$ & $r_c\,(H^{-1}_0)$ & $z_{q=0}$ \\ \hline
Gold Sample   & $0.34^{+0.07}_{-0.08}$ & $0.24^{+0.04}_{-0.04}$
 & $1.02^{+0.09}_{-0.09}$ & $0.78^{+0.24}_{-0.22}$ \\
Gold+BAO      & $0.272^{+0.021}_{-0.019}$ & $0.211^{+0.023}_{-0.027}$
 & $1.089^{+0.070}_{-0.059}$ &  $0.84^{+0.11}_{-0.13}$\\
SNLS          & $0.23^{+0.14}_{-0.17}$ & $0.20^{+0.06}_{-0.07}$
 & $1.12^{+0.20}_{-0.17}$ & $0.91^{+0.66}_{-0.61}$ \\
SNLS+BAO      & $0.265^{+0.019}_{-0.018}$ & $0.216^{+0.013}_{-0.014}$
 & $1.076^{+0.035}_{-0.032}$ & $0.87^{+0.08}_{-0.09}$ \\
Gold+SNLS     & $0.31^{+0.07}_{-0.06}$ & $0.23^{+0.03}_{-0.03}$
 & $1.04^{+0.07}_{-0.07}$ & $0.81^{+0.20}_{-0.22}$\\
Gold+SNLS+BAO & $0.270^{+0.018}_{-0.017}$ & $0.216^{+0.012}_{-0.013}$
 & $1.076^{+0.032}_{-0.030}$ & $0.86^{+0.07}_{-0.08}$\\
\hline
\end{tabular}
}
\end{center}
\end{table}

%
%


\section{Conclusion and discussion}

Observations of SNe Ia indicate that the expansion of the universe is accelerating. 
What drives the acceleration, however, is still acompletely open question. 
From the observational viewpoint, it is of fundamental  importance to differentiate between the two major possibilities, namely,  the existence 
of new fields in high energy physics (dark energy) or modifications of gravitation theory on large scales. 
In this paper, we have focused our attention on one of the leading contender in modified-gravity explanation of acceleration, the so-called 
DGP model. We have analyzed the DGP model by using the gold SN Ia sample,
  the recent SNLS data and the SDSS baryon acoustic oscillations.
Since SN Ia data are sensitive to the value of $\Omega_{r_c}$
  while the baryon acoustic oscillations are sensitive to the
  value of $\Omega_m$, the combination of these data sets breaks the
  degeneracies between the model parameters and leads to strong
  constraints on them, as shown in Figures~2,3,4.
The joint analysis strongly indicates a spatially closed universe,
  which was already obtained by fitting the combination of
  SN Ia data and the X-ray gas mass fraction in galaxy
  clusters (Zhu and Alcaniz 2005; Alcaniz and Zhu 2005).
We also estimate the transition redshift $z_{q=0} \ge 0.70$ at $2\sigma$ 
  confidence level.


In summary, we have discussed the gravitational leakage into extra dimensions as an alternative mechanism for the late-time acceleration 
of the universe (and an alternative route to the dark energy problem). In agreement with other recent analysis, we have shown that a spatially 
closed DGP scenario with a crossover scale $r_c\, \sim H^{-1}_0$ is largely favored by most of the current observational data.


\acknowledgements

This work was partially supported by the National Natural Science Foundation of China, under Grant No. 10533010 and 90403032, 
National Basic Research Program of China under Grant No.  2003CB716300, and by SRF, ROCS, SEM of China. JSA is supported by 
CNPq (Brazilian Agency) under Grants No. 307860/2004-3 and 475835/2004-2 and by Funda\c{c}\~ao de Amparo \`a Pesquisa do 
Estado do Rio de Janeiro (FAPERJ) No. E-26/171.251/2004.

\clearpage
                                                                                
\begin{figure}
\includegraphics[angle=-90, width=16cm]{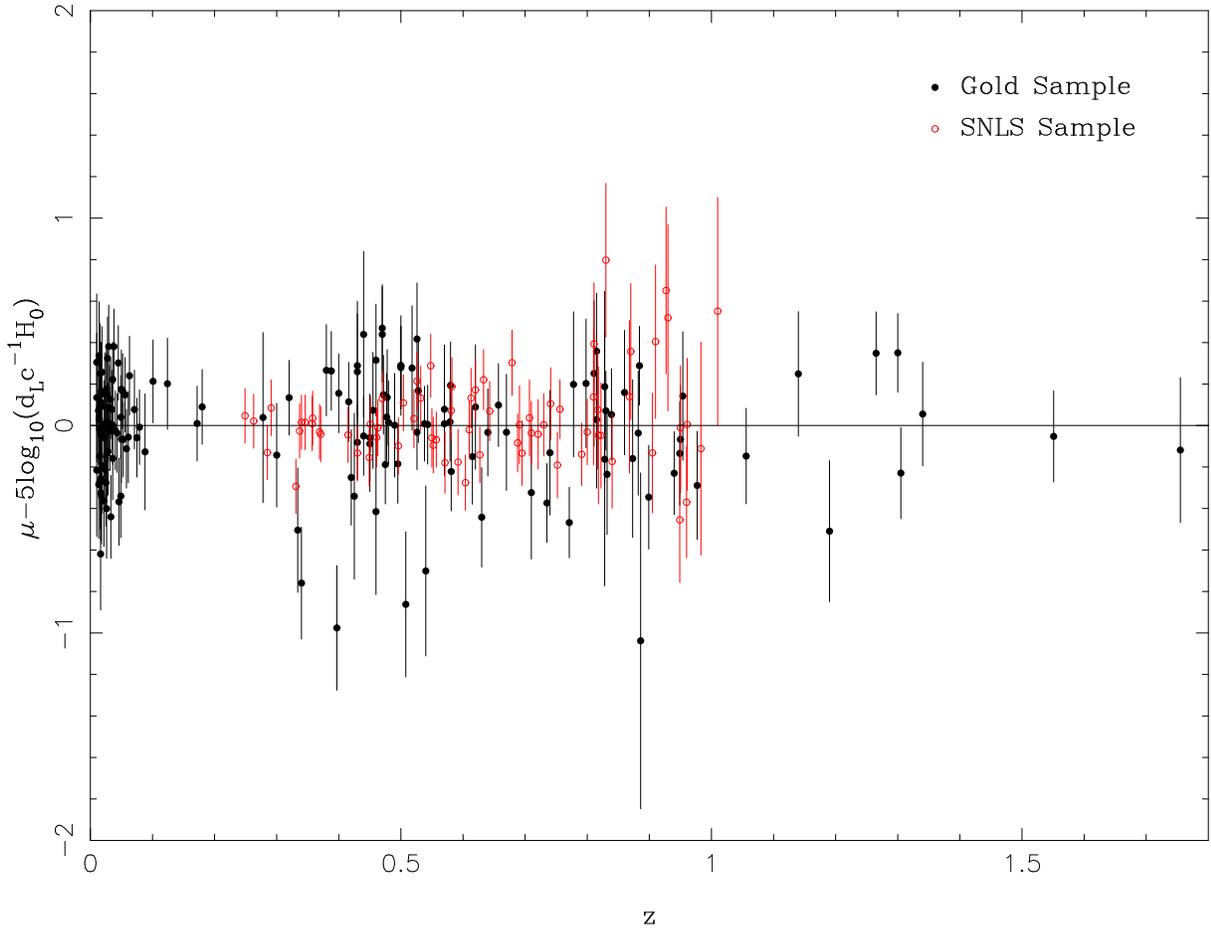}
\figcaption{The gold sample and the SNLS sample are shown in a residual 
	    Hubble diagram with respect to the DGP model with the best-fit 
	    parameters, ($\Omega_m$, $\Omega_{r_c}$) = (0.270, 0.216).
	    \label{Fig_data}
           }
\end{figure}

\clearpage

\begin{figure}
\includegraphics[width=16cm]{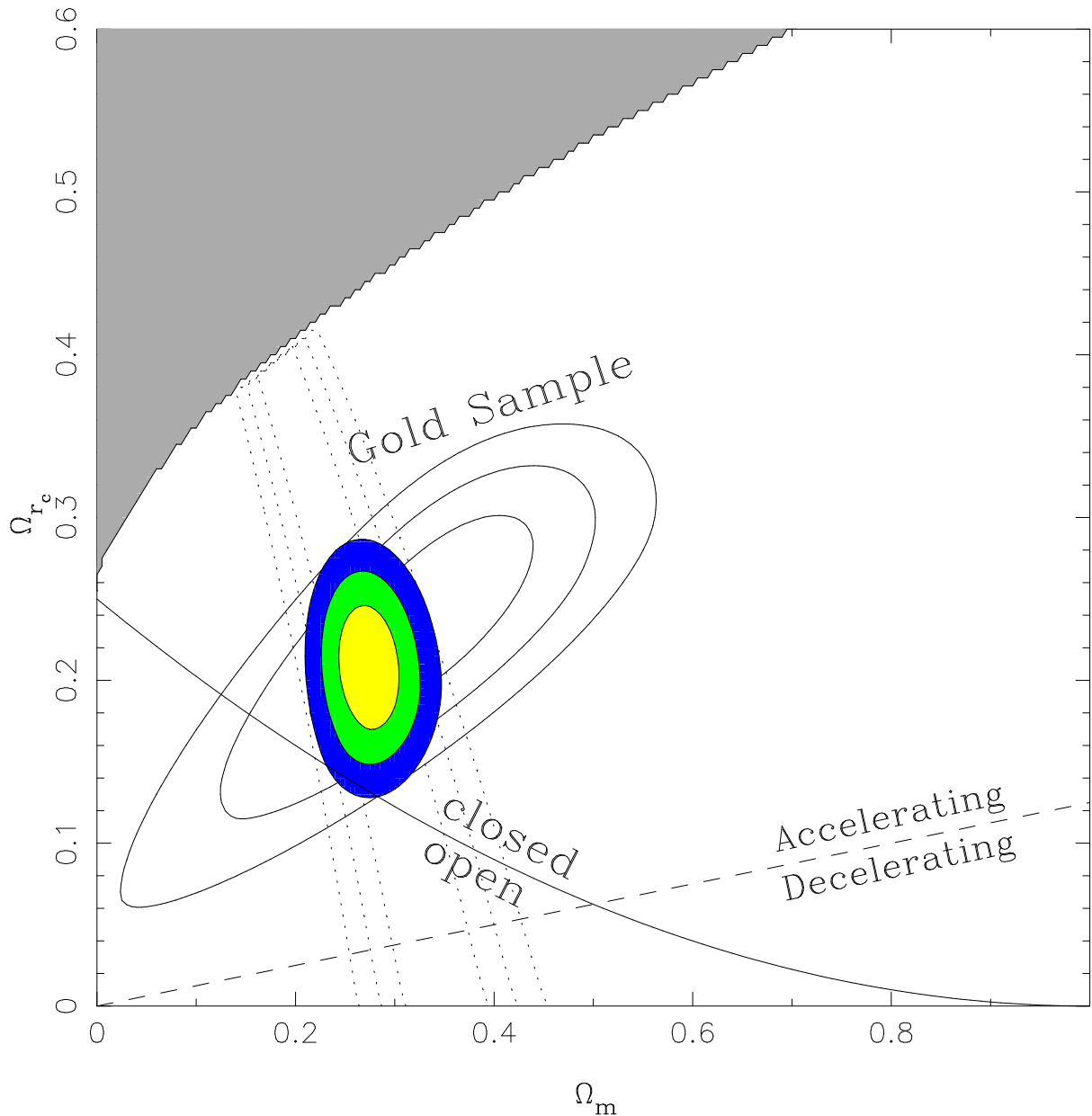}
\figcaption{Probability contours at 68.3\%, 95.4\% and 99.7\% confidence 
	    levels for $\Omega_m$ versus $\Omega_{r_c}$ in the DGP model 
	    from the gold sample of SNeIa data (solid contours), from the
	    baryon acoustic oscillations found in the SDSS data (dotted lines) 
	    and from the combination of the two databases (coloured contours) 
	    -- see the text for a detailed description of the method.
	    The upper-left shaded region represents the ``no-big-bang"
	    region, the thick solid line represents the flat universe
	    and accelerated models of the universe are above the the
	    dashed line.
	    The best fit happens at $\Omega_m=0.272$ and $\Omega_{r_c}=0.211$.
	    \label{Fig_cont13} 
	   }
\end{figure}

\clearpage

\begin{figure}
\includegraphics[width=16cm]{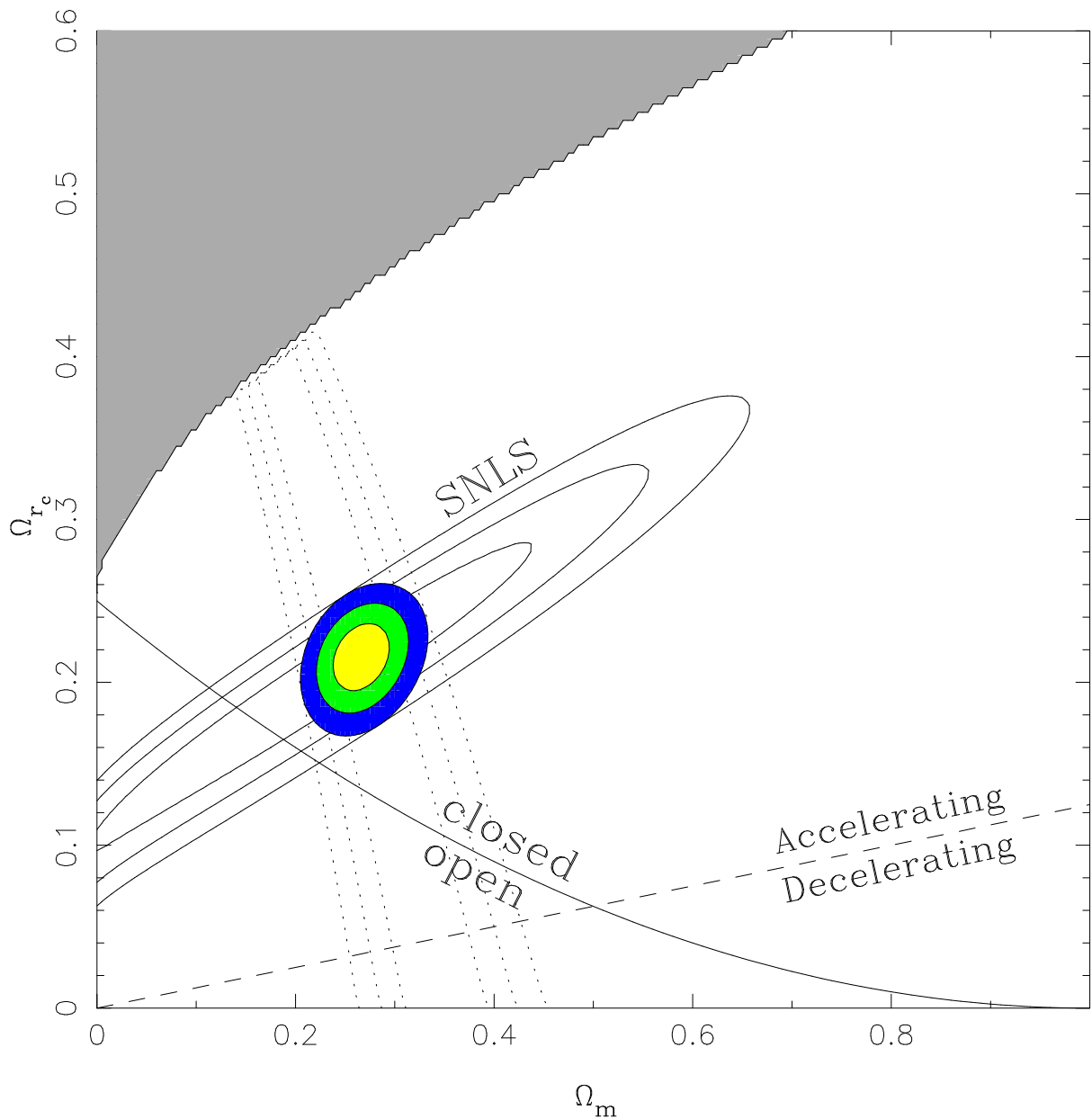}
\figcaption{Probability contours at 68.3\%, 95.4\% and 99.7\% confidence
            levels for $\Omega_m$ versus $\Omega_{r_c}$ in the DGP model
            from the first year SNLS data (solid contours), from the
	    baryon acoustic oscillations found in the SDSS data (dotted lines)
	    and from the combination of the two databases (coloured contours).
	    The upper-left shaded region represents the ``no-big-bang"
            region, the thick solid line represents the flat universe
            and accelerated models of the universe are above the the
            dashed line.
            The best fit happens at $\Omega_m=0.265$ and $\Omega_{r_c}=0.216$.
	    \label{Fig_cont23}
	   }
\end{figure}

\clearpage
                                                                                                                             
\begin{figure}
\includegraphics[width=16cm]{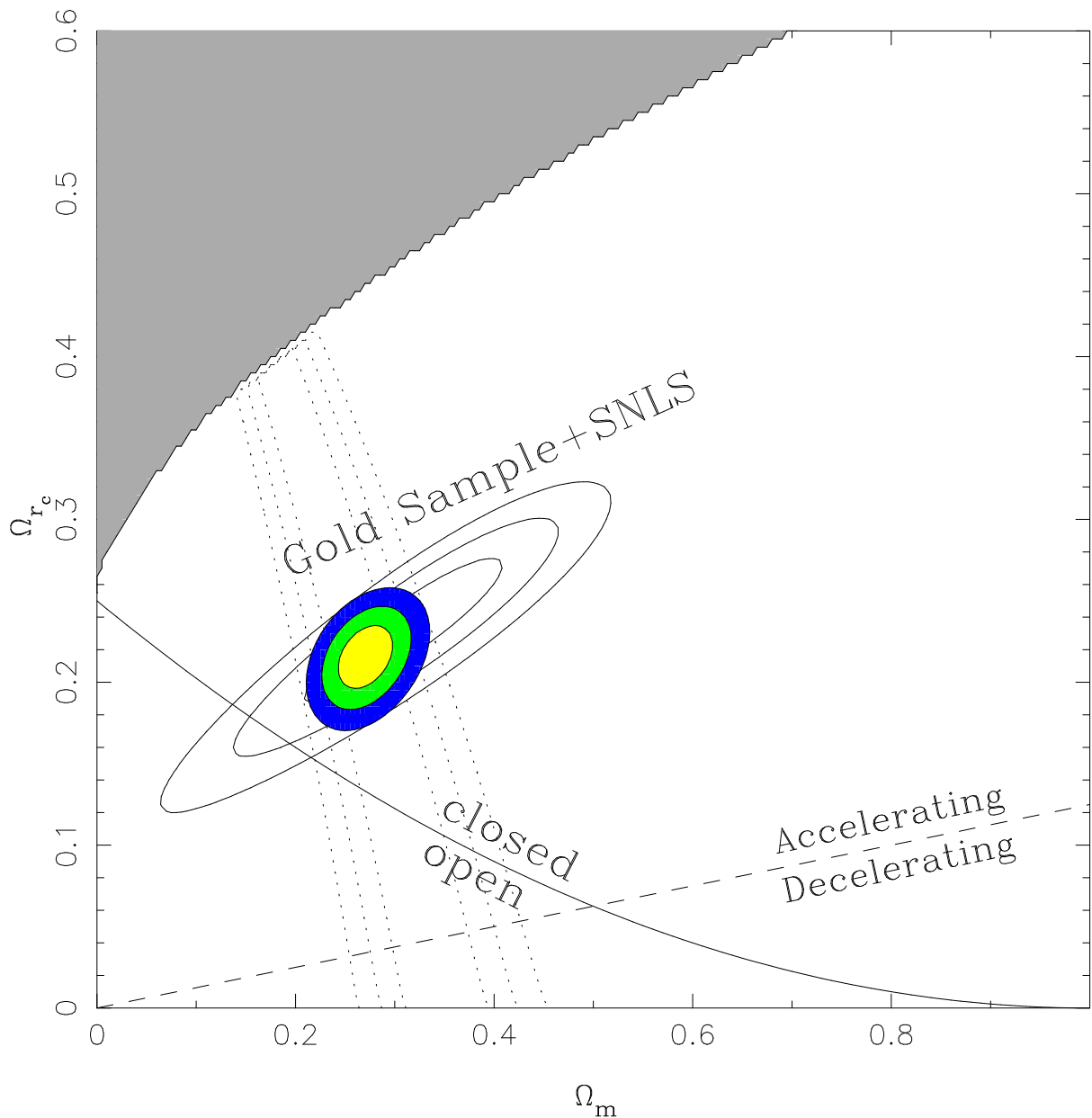}
\figcaption{Probability contours at 68.3\%, 95.4\% and 99.7\% confidence
            levels for $\Omega_m$ versus $\Omega_{r_c}$ in the DGP model
	    from the combination of both the gold sample of SN Ia data and
	    the first year SNLS data (solid contours), from the
            baryon acoustic oscillations found in the SDSS data (dotted lines)
            and from the conjunction of the three databases (coloured contours).
            The upper-left shaded region represents the ``no-big-bang"
            region, the thick solid line represents the flat universe
            and accelerated models of the universe are above the the
            dashed line.
            The best fit happens at $\Omega_m=0.270$ and $\Omega_{r_c}=0.216$.
            \label{Fig_cont123}
           }
\end{figure}

\clearpage

\begin{figure}
\includegraphics[width=16cm]{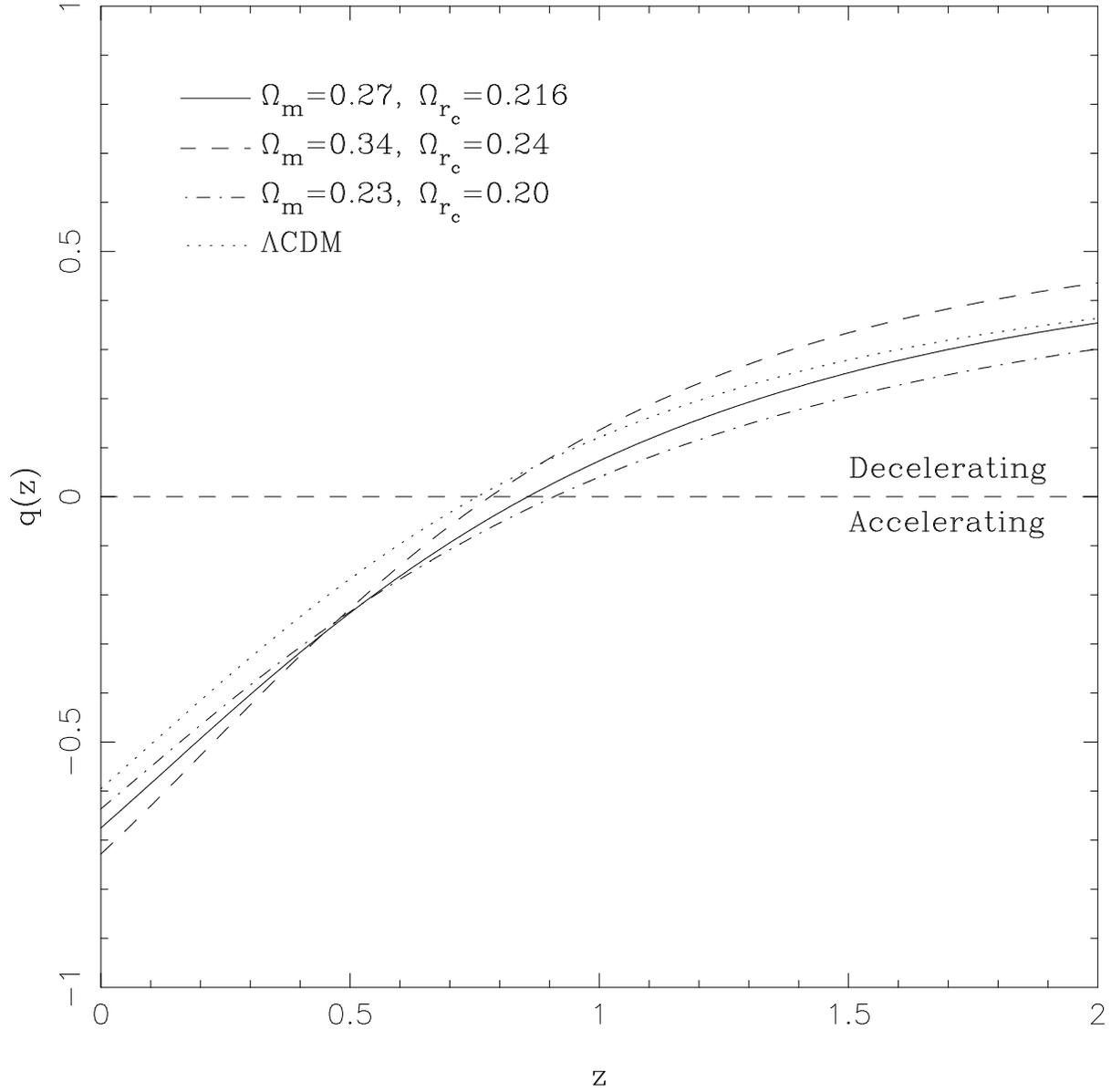}
\figcaption{The deceleration parameter as a function of redshift $z$ for 
	    some best-fit values in DGP model and the standard $\Lambda$CDM.
	    \label{Fig_turnaround}
	   }
\end{figure}
\end{document}